\RequirePackage{fix-cm}
\documentclass[smallextended]{svjour3}  
\UseRawInputEncoding 

\smartqed  % flush right qed marks, e.g. at end of proof
\usepackage{graphicx}
\usepackage{caption}
\usepackage{subcaption}
\usepackage{float}
\usepackage{multirow}
\usepackage{hyperref} 
\usepackage{siunitx}
\usepackage{amssymb}
\usepackage{xr}
\usepackage[table,xcdraw,dvipsnames]{xcolor}
\usepackage{booktabs}
\usepackage{afterpage}
\usepackage{graphicx}
\usepackage{booktabs}
\usepackage{longtable}
\usepackage{tabu}
\usepackage{listings}
\usepackage{hyperref}
\usepackage{tcolorbox}
\usepackage{float}
\usepackage{subcaption}
\usepackage{amsmath}
\usepackage{multirow}
\usepackage[normalem]{ulem}
\usepackage{footnote}
\makesavenoteenv{tabular}
\usepackage{lscape}
\usepackage{array}
\usepackage{enumitem}

\makeatletter
\newcommand*{\addFileDependency}[1]{% argument=file name and extension
  \typeout{(#1)}
  \@addtofilelist{#1}
  \IfFileExists{#1}{}{\typeout{No file #1.}}
}
\makeatother

\begin{document}
\raggedbottom
\title{Domain-based user embedding for competing events on social media %\thanks{Grants or other notes
}
\subtitle{\\}


\author{{Wentao Xu$^1$}
        \and {Kazutoshi Sasahara$^2$}
}


\institute{
\email{myrainbowandsky@gmail.com} \\ \\
$^1$ Department of Science and Technology of Communication, University of Science and Technology of China, China\\
$^2$ School of Environment and Society, Institute of Science Tokyo, Japan
}

\date{Received: date / Accepted: date}

\maketitle

\begin{abstract}
Social divide and polarization have become significant societal issues.
To understand the mechanisms behind these phenomena, social media analysis offers research opportunities in computational social science, where developing effective user embedding methods is essential for subsequent analysis. Traditionally, researchers have used predefined network-based user features (e.g., network size, degree, and centrality measures). However, because such measures may not capture the complex characteristics of social media users, in our study we developed a method for embedding users based on a URL domain co-occurrence network. This approach effectively represents social media users involved in competing events such as political campaigns and public health crises. We assessed the method's performance using binary classification tasks and datasets that covered topics associated with the COVID-19 infodemic, such as QAnon, Biden, and Ivermectin, among Twitter users. Our results revealed that user embeddings generated directly from the retweet network and/or based on language performed below expectations, whereas our domain-based embeddings outperformed those methods while reducing computation time. Therefore, domain-based embedding offers an accessible and effective method for characterizing social media users in competing events.

\end{abstract}

\section{Introduction}\label{sec1}
Social divide and polarization have become significant societal issues. Understanding the mechanisms behind these phenomena is crucial, and social media analysis plays a vital role in this endeavor. 
Competing events have been extensively analyzed on social networking services (SNSs) across various topics, including gender~\cite{Sham2024}, politics~\cite{Petter2022,Hartman2022,Jost2022,Brady2023,Cava2023}, climate change~\cite{Spampatti2023}, COVID-19~\cite{Zhang2021,Aiello2021,Wang2022}, vaccination~\cite{Schmidt2018,Miyazaki2022,Lee2023}, and misinformation~\cite{Shi2019,Allen2020,Grning2024}. 
For our study, we define ``competing events'' as situations involving highly controversial topics in which two distinct user clusters, each with completely opposing viewpoints, can be identified, resulting in polarization. 
Polarization often leads to the formation of echo chambers, where segregated user groups with divergent views reinforce their beliefs within homophilous, like-minded communities~\cite{Kiran2018,Jasny2015}. In these polarized discussions, social networks typically evolve into two opposing groups~\cite{Adamic2005,Nikolov2021,Essig2024}. 

Analyzing user dynamics during these events is crucial for understanding the origins of bias, user engagement, and the strategies influencing polarized clusters on social media. However, there is still a lack of accurate and accessible computational methods---beyond traditional ones---to better identify and characterize opposing users for computational social science purposes. A possible method of quantifying competing events involves embedding techniques, which projects complex data into low-dimensional vector spaces derived from high-dimensional or unstructured data.
For example, Word2vec~\cite{1301.3781} is a well-known word-embedding method widely used in ``text-as-data'' research involving text classification~\cite{lilleberg2015support}, cluster analysis~\cite{Austin2019}, and the study of the evolution of topics~\cite{10.1016/j.ipm.2019.02.014} in the social sciences.
Beyond textual data, embeddings can also be generated from the latent features of node relationships within graphs~\cite{DBLP:journals/debu/HamiltonYL17,8294302,wang2019,GOYAL201878,10.1145/3483595,Wang2024} or networks~\cite{10.1145/2736277.2741093,cong2019textual,10.1145/3491206,Babul2024}.
These techniques provide a powerful framework for analyzing the dynamic interactions and structural properties of data across various domains.

Building upon these advancements, user embedding has emerged as a vital concept for representing users based on their personal and network attributes, as well as profiling users and predicting their behaviors. Applications of user embedding include quantifying organizational polarization~\cite{Waller2021}, measuring ideological distances~\cite{alatawi2023quantifying}, and identifying opportunities for civic engagement~\cite{deVries2023}. User embedding offers deep insights into user dynamics and patterns of interaction on social media platforms. It also supports a range of downstream tasks, such as fake news detection, fraud detection, and hate speech detection.

Despite the progress made in user embedding, the lack of accurate and accessible methods for characterizing users in the context of competing events on social media limits our understanding of key dynamics, such as echo chambers~\cite{Sasahara2021-tf} and dualistic antagonism~\cite{xu2025}. 
Although there is a strong demand for such computational methods tailored to competing events, existing network-based, content-based, and hybrid methods fall short in classification speed, accuracy, and subsequent analytical utility.
These limitations have motivated us to undertake the current study.
To bridge this methodological gap, we propose a domain-based embedding approach that leverages URL co-occurrence patterns in information sharing. 
The core idea is that shared sources of information, or ``domains,'' can reflect users' shared interests and preferences~\cite{Adamic2005,Pennycook2021}, as similar users tend to share similar content and this content is often linked via URLs linked to online articles, images, and videos.
By exploiting latent features from the domain co-occurrence network, our approach aims to gain deeper insights into competing clusters and provide a more effective and computationally efficient method for characterizing users in the context of competing events on social media. 
To this end, we formulated our research questions as follows:

\begin{enumerate}[leftmargin=1cm]
\item[RQ1:] Does domain co-occurrence in information sharing serve as an effective signal for identifying user clusters in competing events?
\item[RQ2:] To what extent does domain-based user embedding improve performance compared to existing methods?
\end{enumerate}

To answer these questions, we benchmarked our proposed model against various alternatives, including one that employs text-based user embeddings from social media-centric measures~\cite{Islam_Goldwasser_2021}. Our method makes two original and important contributions. First, the domain-based user-embedding technique not only outperforms existing methods in analyzing competing social events but also reduces computational complexity and processing time. Second, it offers the unique capability of measuring users' similarity, an area in which traditional methods of community detection fall short. We expect our model to be particularly effective for user classification and similarity measurement in events characterized by intense competing user dynamics, such as major presidential elections. By providing insights into these events, our approach can help illuminate the mechanisms behind social divide and polarization, contributing to the development of countermeasures to mitigate their negative social impacts.

The rest of the paper is organized as follows. The Related Work section reviews two types of embedding methods, which are the basis of our method and benchmark models. 
The Data and Methods section describes the dataset, the three topics analyzed, our proposed model, benchmark models, and evaluation methods.
Parameter settings for the proposed and baseline models are also presented. 
The Results section describes the results from performance comparisons and the application of the proposed model. 
The Discussion section addresses implications and limitations of our method, and the Conclusion section summarizes our findings. 
Finally, the Future Directions section outlines future research directions for user embedding in the context of competing events.

\section{Related Work}
This section reviews network and content embedding methods, which are fundamental to our model and the benchmark models. Our model relies solely on network embedding, whereas the other models make use of network embedding, content embedding, or a combination of both.

\subsection{Network Embedding}
Node embedding is a technique used to transform a large number of nodes in a network into a low-dimensional space where their structural properties are preserved.
Node embeddings are typically used in three activities: node prediction and clustering, node visualization, and link prediction.
DeepWalk~\cite{10.1145/2623330.2623732} is a Skip-gram~\cite{1301.3781}-based node embedding approach that, based on random walks, learns latent representations of nodes in a network. 
It begins by generating short random walks, each of which is treated as a sequence of words in a sentence. 
It subsequently uses the Skip-gram model in Word2vec to discover a node's latent representation. 

Node2Vec~\cite{node2vec} is a variant of DeepWalk that learns node representations using the Skip-gram model, based on a collection of nearby nodes generated through biased random walks. 
It extends DeepWalk by introducing a biased random walk mechanism that interpolates between breadth-first search (BFS) and depth-first search (DFS).
This strategy balances the embedding between homophily and structural equivalence.
As a result, Node2Vec can more effectively preserve both second-order and higher-order proximity.
The resulting co-occurrence network can reveal similarities in the connections among users, words, and hashtags.

The factorization-based approach offers an alternative for node embedding, as exemplified by techniques such as Graph Factorization~\cite{10.1145/2488388.2488393}, GraRep \cite{10.1145/2806416.2806512}, and HOPE \cite{10.1145/2939672.2939751}.
The methodology is intuitively simple: learn embeddings for each node such that the inner product of the acquired embedding vectors approximates a deterministic measure of node similarity.
GraRep has been shown to outperforms LINE \cite{10.1145/2736277.2741093} and DeepWalk in the clustering and classification of node embedding in the BlogCatalog network \cite{nr} and DBLP network \cite{Tang:08KDD}.
\cite{10.1145/2939672.2939751} have significantly contributed to the reconstruction, link prediction, and node recommendation tasks by estimating high-order proximities, while \cite{Sun2018MultiviewNE} have proposed graph factorization network embedding based on multiview clustering. 
This multiview-based factorization approach demonstrated performance comparable to Node2vec, DeepWalk, and LINE while testing on BlogCatalog, PPI \cite{10.1093_nar_gku1204}, and Wikipedia \cite{wikidata}.

Structural Deep Network Embedding (SDNE)~\cite{10.1145/2939672.2939753} is another approach to node embedding that uses first-order proximity to preserve local similarity while concurrently learning both local and global structures of a hierarchy. 
Around the same time, deep neural graph representation (DNGR)~\cite{cao2016deep} was also proposed; they encode and decode the positive the pointwise mutual information (PPMI) matrix using a stacked denoising autoencoder and multilayer perceptrons. 

Another class of node embedding methods is graph neural networks (GNN) and geometric deep learning. 
Among the various forms of GNNs, encoders based on graph convolutional neural network (GCN) \cite{kipf2016semi} are some of the most widely used for unsupervised learning.
A GCN learns hidden layer representations that encode both the local graph structure and features of nodes; each node in a network is represented by a vector based on the node feature matrix and adjacency matrix. 
%Finally, GraphSage~\cite{1706.02216}---a spatial convolution method---is an inductive framework that leverages the textual information of vertices to generate network representations by sampling and aggregating features from a node's local neighbors. 

Table~\ref{tab:embeddings} summarizes common network embedding methods.

%Table 1
\begin{table}[t]
\caption{\label{tab:embeddings} Network embedding methods, categorized by the source of node samples and
method for representation learning.}
\begin{tabular}{lll}
\hline Method & Source of node samples & Model \\
\hline DeepWalk & Truncated random walks hierarchical softmax & Skip-gram  \\
LINE & First-and second- order proximity & Skip-gram \\
Node2vec & Biased Truncated Random Walks & Skip-gram \\
GraRep & $A^{i}$ ,where $i=1,2, \cdots, k$, & Matrix Factorization \\
     &  (i.e., a k-step probability transition matrix) &  \\
SDNE &  First-and second- order proximity & Deep Autoencoder \\
DNGR & Random surfing & Stacked Denoising Autoencoder \\
\hline
\end{tabular}
\end{table}

Unsupervised node embedding algorithms have been extensively studied and generally aim to learn both local and global latent structures to represent nodes in a network. Although no single method can be universally recommended due to task specificity, recent research on complex network embedding~\cite{DehghanKooshkghazi2022} argued that ``If one needs to pick one (node) embedding algorithm before running the experiments, then Node2vec is the best choice as it performed best in our tests.''
We therefore used Node2vec to compute node embeddings both in our proposed model and benchmark models.

\subsection{Content Embedding}
Along with network embedding, content and/or text features are often used to generate user embeddings. 
We also tested such models by comparing them with our proposed model, described later. 
BERT, a well-known natural language processing algorithm, was introduced by \cite{1810.04805} as a pretrained language model (LM).
It uses a transformer architecture~\cite{1706.03762} trained for both a masked LM and a next-sentence prediction task. A masked LM predicts randomly masked words following conditioning on both their left and right contexts. The attention mechanism in the transformer, based on an encoder-decoder structure, enables BERT to analyze each word in a sentence in relation to all other words. 

This mechanism underlies a key distinction between BERT and Word2vec: BERT supports polysemy disambiguation, allowing it to generate different vectors for a single word depending on context. In contrast, Word2vec assigns a fixed vector to each word, regardless of context, by retrieving it from a lookup table. BERT inherits this context-sensitivity from ELMo~\cite{1802.05365}, which uses a bidirectional LSTM to compute dynamic embeddings based on the entire sentence.

However, BERT is limited to sequences of up to 512 tokens, which may not suffice for longer texts such as news articles. To overcome this, we adopted Longformer~\cite{beltagy2020longformer}, a transformer variant based on RoBERTa~\cite{1907.11692}, as the primary component of our benchmark model. Longformer supports inputs of up to 4,096 tokens, enabling us to process entire retweeted articles in a single pass.
It employs a sliding window-based attention mechanism for local context and supplements this with global attention applied to selected input tokens. Due to its stacked layers, Longformer achieves a broad receptive field, similar to that of stacked convolutional neural networks (CNNs)\cite{wu2019pay}. Following\cite{beltagy2020longformer}, we configured the global attention to be symmetric: tokens with global attention attend to all other tokens in the sequence, and all tokens attend to them. This hybrid attention mechanism reduces computational complexity while maintaining high attention performance.

\section{Data and Methods}

\subsection{Data}\label{section:datasets}
We repurposed the COVID-19 tweet dataset from our previous research~\cite{Xu2021,Xu2022}. 
We used Twitter (now called ``X'') to collect social data for our study. 
Over a 24-month period between February 20, 2020 and February 28, 2022, we used the Twitter Search API to gather tweets (now called ``posts'') by querying COVID-19-related keywords: ``corona virus,'' ``coronavirus,'' ``covid19,'' ``2019-nCoV,'' ``SARS-CoV-2,'' and ``wuhanpneumonia.''~\footnote{The X Pro API is now required to collect tweets (posts) on X; the free and academic APIs can no longer be used for that purpose.}
User properties in the dataset, such as bot scores, were obtained from previously published research~\cite{Xu2021,Xu2022}.
Retweet (now called ``repost'') data were used to construct retweet networks, as explained later. 
A bot score is the likelihood that a user is a social bot, calculated by Botometer v4~\cite{Sayyadiharikandeh_2020}.
In this study, we refer to this dataset as the ``COVID-19 tweet dataset,'' and used it to evaluate the performance of our proposed method alongside other benchmarks. We focused on three topics as competing events: ``QAnon,'' a conspiracy theory topic; ``Biden,'' a political topic; and ``Ivermectin,''  a topic related to COVID-19. 
%Table~\ref{tab:data} provides a summary of these topic-related tweets.

\subsection{Data Pre-processing}
Two of the benchmark models used in our study incorporate linguistic features for constructing user embeddings. In this purpose, preprocessing is essential to saving computing time and resources and increasing accuracy. All text data were converted to lowercase, as language models are case-sensitive, and email addresses, punctuation, numbers, currency symbols, and the ``re'' prefix in retweet objects were removed to improve accuracy. Hashtags were also removed to eliminate the association between hashtags and text content. To compare the latent features of texts and retweeted online articles for the benchmark models, only tweets with non-null values in both text objects and retweeted objects were maintained. Consequently, the sizes of the training datasets were 274, 1,000, and 582 for the topics of QAnon, Biden, and Ivermectin, respectively (Table~\ref{tab:training}).
These preprocessed data were subsequently fed into the Longformer-based model, which was used as a benchmark for comparison.

%Table 2
\begin{table}[h]
\caption{Number of users/nodes in the training dataset}
\label{tab:training}
\begin{tabular}{cccccc}
\hline
Topic      & \multicolumn{1}{l}{\begin{tabular}[c]{@{}l@{}}Domain-based\\ user embedding\end{tabular}} & \begin{tabular}[c]{@{}c@{}}Hashtag-based \\ user embedding\end{tabular} & Text-based  & \begin{tabular}[c]{@{}c@{}}Retweeted\\  articles\end{tabular} & Retweet networks   \\ \hline
QAnon      & 292                                                                                       & 794                                                                       & 1,000 & 274                                                           & 1,000 
\\
Biden      & 1,000                                                                                     & 1,000                                                                     & 1,000 & 1,000                                                         & 1,000 \\ 
Ivermectin & 312                                                                                       & 18                                                                        & 578   & 582                                                           & 1,000 \\ \hline
\end{tabular}
\end{table}

\subsection{Louvain Clustering as Reference}\label{subsec:louvain}
\label{labeling}
When it comes to controversial topics, social ties are often split between supporters and opponents, a result of which is echo chambers~\cite{Sasahara2021-tf}.
Those divisions are prime examples of competing user clusters, which are commonly identified using the Louvain algorithm~\cite{blondel2008fast}, a modularity-based community detection method. In our study, we used the results of Louvain clustering as references, namely binary labels for two competing communities---pro-$X$ and anti-$X$---in which X represents the topics QAnon, Biden, or Ivermectin in order to evaluate the embedding results. To that end, we used the retweet network as the target to which we applied the Louvain algorithm. 

Retweeting (i.e., ``reposting'') is an important information-sharing behavior by which followers can receive messages from their favorite influencers immediately. A retweet can be both a productive, communicative tool and a selfish act of an attention seeker ~\cite{Boyd2010TweetTR}. 
\cite{Suh2010,10.1145/1772690.1772751,Hong2011} have argued that a user's contextual information (e.g., social network topology, tweet content, and URL) affects retweeting behavior. Therefore, a retweet network is used to label users in competing clusters.

\begin{figure}[p]
\begin{center}
\includegraphics[width=\linewidth]{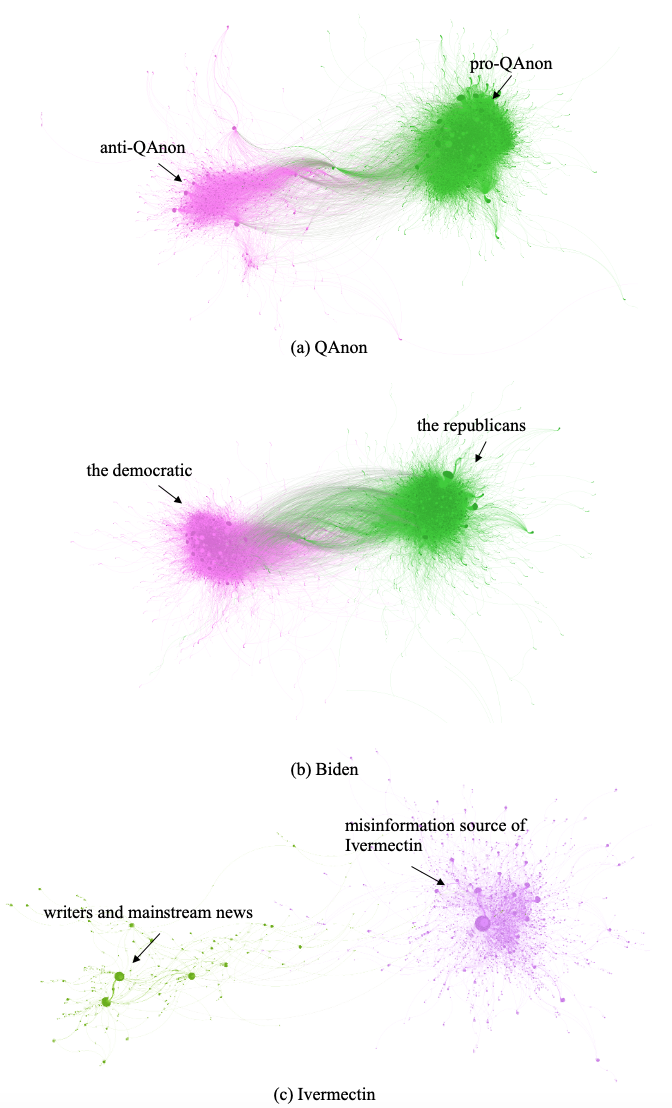}
\end{center}
\caption{Retweet networks of three topics: (a) QAnon, in which green represents pro-QAnon users and magenta represents anti-QAnon users; (b) Biden, in which green represents Republicans and magenta represents Democrats; and (c) Ivermectin, in which green represents writers and mainstream news and magenta represents users diffusing misinformation about Ivermectin.}
\label{fig:RT}
\end{figure}

We relied on the data and user labels provided in our published papers~\cite{Xu2021,Xu2022}, in which user properties involved in the competing events are described. As mentioned, the Louvain algorithm was applied to identify competing clusters of users in the retweet networks using the network analysis software Gephi~\cite{Jacomy2014}~\footnote{\url{https://gephi.org/}}. 
Competing users were harvested from the retweet network when they related to the topics of QAnon, Biden, or Ivermectin (Figure~\ref{fig:RT}).
Users in each topic were segregated---that is pro-QAnon and anti-QAnon in Figure~\ref{fig:RT}(a), Democratic and Republican users as in Figure~\ref{fig:RT}(b), and Ivermectin-related misinformation and mainstream media users as in Figure~\ref{fig:RT}(c). 
The user-related statistics of the three topics appear in Table~\ref{tab:data} and \ref{tab:nodes}.
%Second, we annotated the top two giant clusters as two competing clusters.
To obtain the semantic implications of user clusters, the profiles of the top three largest in-degree users (i.e.,　users with many followers) were examined for each topic. In that way, two classes of competing users were identified for each topic.

%Figure 1

% Table 3
\begin{table}[t]
\caption{Size of retweet networks across three topics}
\centering
\begin{tabular}{llll}
\hline
Topic      & \#nodes & \#edges & keywords          \\ \hline
QAnon      & 141,334  & 259,327  & QAnon, deep state \\ 
Biden      & 29,0504  & 503,489  & Biden, Joebiden   \\ 
Ivermectin & 23,047   & 24,993   & Ivermectin        \\ \hline
\end{tabular}
\label{tab:data}
\end{table}

%Table 4
\begin{table}[t]
\centering
\caption{Size of competing clusters across topics}
\label{tab:nodes}
\begin{tabular}{ccc}
\hline
Topic      & Cluster 1   & Cluster 2               \\ \hline
QAnon      & 103,305 (pro-QAnon)       & 32,179 (anti-QAnon)                   \\
Biden      & 104,504 (the democratic) & 88,079 (the republicans)             \\ 
Ivermectin & 10,727 (Misinformation)  & 3,582 (Mainstream media and writers) \\ 
\hline
\end{tabular}
\end{table}

\subsection{Domain-based User Embedding}\label{section:domain-based}
We propose a method called ``domain-based user embedding,'' which focuses on co-occurrence patterns of URLs in social media shares, including retweets (reposts) on platforms such as Twitter (X). The method is predicated on two assumptions: first, that similar users tend to share similar content, and second, that sharing content often involves sharing URLs linked to online materials such as articles, images, and videos on the Internet. Based on those premises, we hypothesized that shared sources of information, or ``domains,'' can reflect users' shared interests and preferences. We leveraged that behavioral characteristic to enhance our understanding of user dynamics, in an approach supported by empirical evidence. Social media studies have frequently revealed co-occurrence similarity in statistical connections~\cite{BARNETT201738,said2019mining}, and indicated that it can effectively quantify associations between messages~\cite{e24020174}. 
Domains have also been used to evaluate sources of information~\cite{Pennycook2021}. 
Therefore, embedding latent features based on domain co-occurrences may accurately reflect a user's interests, positions, and/or stances.

\begin{figure}[ht]
\begin{center}
\includegraphics[width=0.7\linewidth]{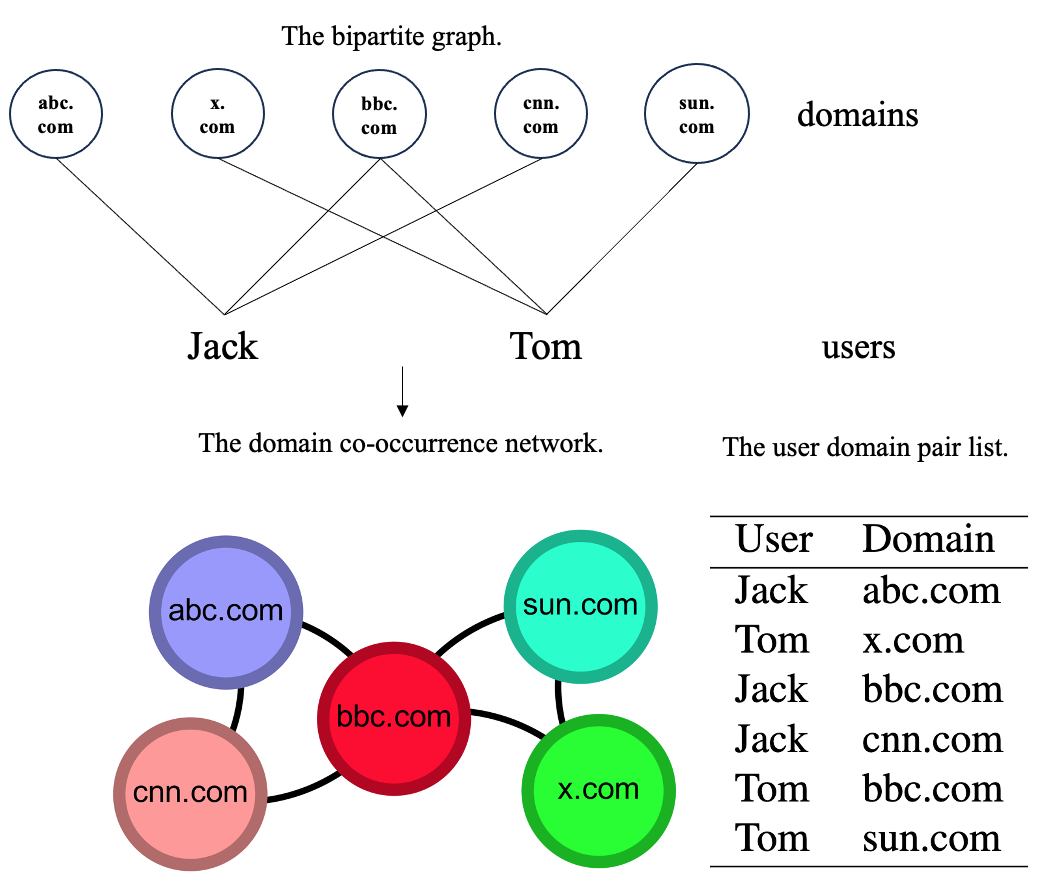}
\end{center}
\caption{Example of a domain co-occurrence network. From a given list of domain co-occurrences, a bipartite graph is constructed (top) and subsequently projected onto a domain co-occurrence network. The figure is for demonstrative purposes only.} 
\label{fig:example}
\end{figure}

In our study, domains were extracted from URLs in each retweet. Shortened URLs were expanded to their original forms, from which domains were extracted. For instance, from the URL \url{xxx.rt.com} the extracted domain was \url{rt.com}, which is the website of RT, formerly called ``Russia Today''.  Some domains could not be collected when URLs were inactive or had expired. Two domains were regarded as a co-occurrence if the same user retweeted (reposted) both of them in a single post or distant posts. The domain co-occurrences of two users are illustrated in the list in Figure~\ref{fig:example}.
From those data, a bipartite graph with domains and users was constructed (Figure~\ref{fig:example}, top), which was subsequently projected onto the domains to create a domain co-occurrence network (Figure~\ref{fig:example}, bottom). 
Using the Node2vec algorithm, the embedding representation of each domain was calculated, which allowed each domain (node) to be represented as a vector ($x$). Subsequently, the embedding for each user was computed by summing the vectors of the domains that they were associated with.

Formally, the user embedding $\vec{U}$ is computed as the summation of the vectors of domains (domain embeddings) involved in $U$'s retweets as shown in Equations (\ref{eq:u}). 
\begin{equation}
\label{eq:u}
  \vec{U}=\sum_{i=1}^{n} \vec{x_i}, 
\end{equation}
where ${x_i}$ is the $i$-th domain embedding associated with user $U$.

For example, user embeddings for Jack and Tom in Figure~\ref{fig:example} are calculated in Equation (\ref{eq:1}) and (\ref{eq:2}):

\begin{align}
\label{eq:1}
    \vec{Jack} &= \vec{abc.com} + \vec{bbc.com} + \vec{cnn.com} \\
\label{eq:2}
    \vec{Tom} &= \vec{x.com} + \vec{bbc.com} + \vec{sun.com}
\end{align}

%Figure 2

\begin{figure}[h]
\centering
\includegraphics[width=0.7\linewidth]{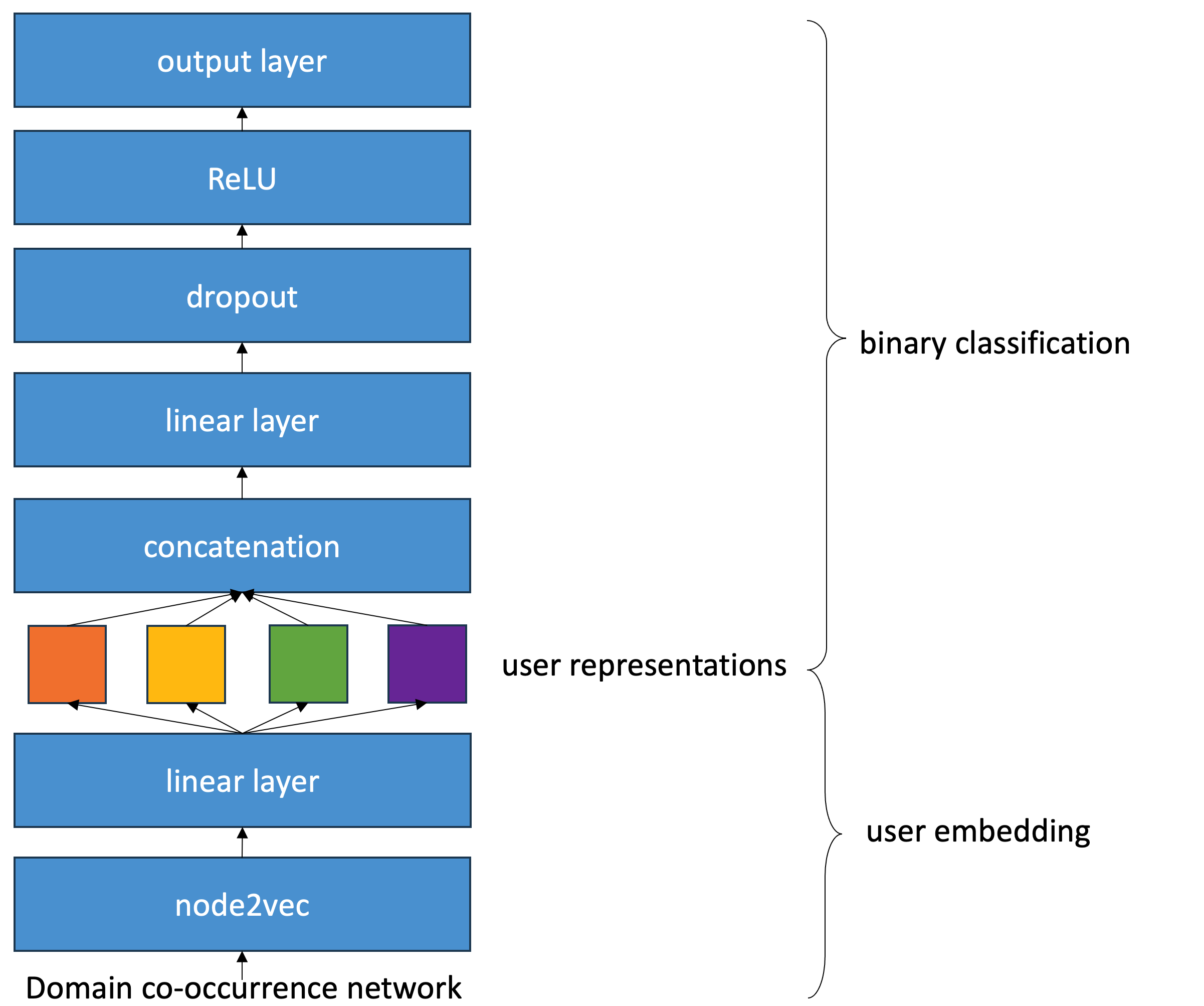}
\caption{Architecture and procedures of the proposed model. A domain co-occurrence network is constructed with the mechanism exemplified in Figure 2. The Node2vec embedding for each user is calculated and input into the liner layer of a neural network, while the user representations are concatenated and sent into another linear layer. The dropout layer is used to reduce overfitting, and ReLU, used as an activation function, introduces nonlinearity. The final classification layer is obtained through the output layer.}  
\label{fig:architechture}
\end{figure}

The process of our proposed model consists of three steps: constructing a domain co-occurrence network, deriving user embeddings from the network, and comparing the results of binary classifications with those of Louvain clustering, which serves as the reference (Figure~\ref{fig:architechture}).
In Step 1, the undirected and unweighted domain co-occurrence network was constructed using the Python library NetworkX, and each user was labeled based on the Louvain method~\cite{blondel2008fast} (see Section~\ref{labeling}). 
For each user in the training dataset, we used Node2vec (i.e., fastnode2vec library)~\footnote{https://github.com/louisabraham/fastnode2vec} to compute user embedding in Step 2 and feedforwarded them to a linear layer to compute user representations. The user representations were subsequently concatenated and passed through a two-layer classifier, between which the second linear layer with the ReLU activation function. The dropout technique was used in the architecture to prevent overfitting between connected network layers. The stochastic gradient descent over shuffled mini-batches with the Adam learning technique and cross-entropy objective function were used for binary classification. The details and methods of Step 3 are outlined in Section~\ref{sec:benchmark} and \ref{metric}.

Instead of using a joint-feature model~\cite{Islam_Goldwasser_2021}, which is presumed to be superior to a single-feature approach, our proposed model employed a single feature for user representation; that is, domain co-occurrence patterns. That simplification significantly reduced computational costs while enhancing the detection of competing users on social media.

%Figure3

\subsection{Benchmark Models}\label{sec:benchmark}
For comparisons, we used four benchmark models, each of which utilizes a different type of embedding representation: retweet networks, tweet texts, retweeted online articles, or hashtags~\cite{Xu2022}. 

To construct network-based benchmark models that use retweet network and hashtag co-occurrence network structures independently, we employed Node2vec to obtain embedding representations. Node2vec focuses solely on network topology and disregards other node attributes. This makes it easy to scale to large graphs using techniques like optimization, parallel processing, and efficient data structures, making it useful for many real-world applications. The embedding representation from a retweet network is obtained by applying the Node2vec to its graph object. 

A hashtag co-occurrence network can be another method to characterize topics~\cite{Xu2022}. 
We therefore employed concatenated hashtag-based user embeddings as a benchmark to represent each user involved in competing events. In the model, hashtags in the text object were regarded as a co-occurrence if the same user included them in any of their tweets. For instance, if a user used ``\#QAnon'' ``\#COVID19'' in a tweet and later used ``\#Biden'' and ``\#COVID19''  in another tweet, then, ``\#QAnon'', ``\#COVID19'',and``\#Biden'' would be regarded as co-occurring hashtags for that user.

For content-based benchmark models that use tweet texts and retweeted online articles independently, a ``longformer-base-4096'' model started from the RoBERTa \cite{1907.11692} checkpoint and pretrained on long documents is used for calculating embeddings. The content-based embedding is subsequently forwarded to stacked bidirectional LSTMs. Concatenating the forward and backward directions yields the latent representation of tweets. Such benchmark models employ a context-aware attention technique to assign significant words to the final linguistic representation~\cite{DBLP:journals/corr/BahdanauCB14}.

In sum, we generated user embeddings using the proposed method and compared them with those derived from benchmark models: retweet network embeddings, hashtag co-occurrence network-based embeddings, and content-based embeddings from text and retweeted online articles. 
%To thoroughly evaluate our method, we selected three competing topics of varying sizes: QAnon as a conspiracy theory topic, Biden as a political topic, and Ivermectin as a medical topic. Last, we assessed the performance of those embeddings by comparing the binary classification results with those from Louvain clustering.

\subsection{Evaluation}\label{metric}
To assess the proposed model's effectiveness, we conducted experiments across the topics using the same users, with consistent random states set for all corresponding experiments. For the embedding process, we configured the hidden dimensions for the three topics at 600 and set the dropout rate at 0.5; doing so ensured that the user embeddings produced by our models were uniform in length and equally effective in mitigating overfitting. For network benchmarks such as retweet (RT) and hashtag network embeddings, the dimensions were set at 300, walk length at 100, window size at 10, width (p) at 2.0, and depth (q) at 0.5 for Node2vec training. Parameters for the content-based benchmarks were established following the recommendations of \cite{Islam_Goldwasser_2021}. The sample size for those experiments is detailed in Table~\ref{tab:training}. The dataset was divided with 80\% for training, 10\% for validation, and 10\% for testing. 
%All experiments were carried out on a DELL Precision 7920 Tower running Ubuntu 22.04, equipped with dual Intel(R) Xeon(R) Gold 6136 CPUs and 512GB of RAM. 
The model parameters are summarized in Table~\ref{tab:paras}.

%Table 5
\begin{landscape}
\begin{table}[h]

\centering
\caption{Parameter settings}
\begin{tabular}{cccc}
\hline
\multicolumn{1}{l}{} & \begin{tabular}[c]{@{}c@{}}Content embedding\\ (Tweet / Retweet / Retweeted articles)\end{tabular}                                                                                          & Retweet networks                                                                                                                           & Domain-based/hashtag-based                                                                                                                                                                                                                                                                                                                         \\ \hline
Model                & RoBERT                                                                                                                                                                                      & Node2vec                                                                                                                                   & Our proposed model (Domain-based)                                                                                                                                                                                                                                                                                                                                 \\ \hline
Parameters           & \multicolumn{1}{l}{\begin{tabular}[c]{@{}l@{}}1st linear layer dimensions: 300, 600\\ 2nd linear dimension: 600, 3\\ Number of layers: 2\\ Bidirectional: True\\ Dropout: 0.5\end{tabular}} & \multicolumn{1}{l}{\begin{tabular}[c]{@{}l@{}}Dimension: 300\\ Walk length: 100\\ Window size : 10\\ Width: 2.0\\ Depth: 0.5\end{tabular}} & \multicolumn{1}{l}{\begin{tabular}[c]{@{}l@{}}Node2vec training:\\ Dimension: 300\\ Walk length: 100\\ Window size : 10\\ Width: 2.0\\ Depth: 0.5\\ Batch size = 18\\ \\ Node classification training:\\ 1st linear layer dimensions: 150, 600\\ 2nd linear dimension: 600, 3\\ Number of layers: 2\\ Bidirectional: True\\ Dropout: 0.5\end{tabular}} \\ \hline
\end{tabular}
\label{tab:paras}
\end{table}
\end{landscape}

We evaluated the models using five key metrics: accuracy, F1 score, precision, recall, and area under the receiver operating characteristic curve (ROC-AUC), calculated using Python's scikit-learn.\footnote{\url{https://scikit-learn.org/stable/}}. Regarding the topic QAnon, for example, true positive instances are those in which the model correctly identifies pro-QAnon instances as positive, thereby showcasing its ability to accurately classify nodes with positive labels. True negative instances occur when the model accurately recognizes anti-QAnon instances as being negative, which demonstrates its precision. In contrast, false positive instances, in which anti-QAnon instances are incorrectly labeled as pro-QAnon, highlight the model's susceptibility to erroneous positive classifications. Similarly, false negative instances, in which pro-QAnon instances are mislabeled as anti-QAnon, reveal challenges in capturing key patterns.

\section{Results}
\subsection{Performance comparison}
\begin{figure}[thb]
\begin{center}
\includegraphics[width=\linewidth]{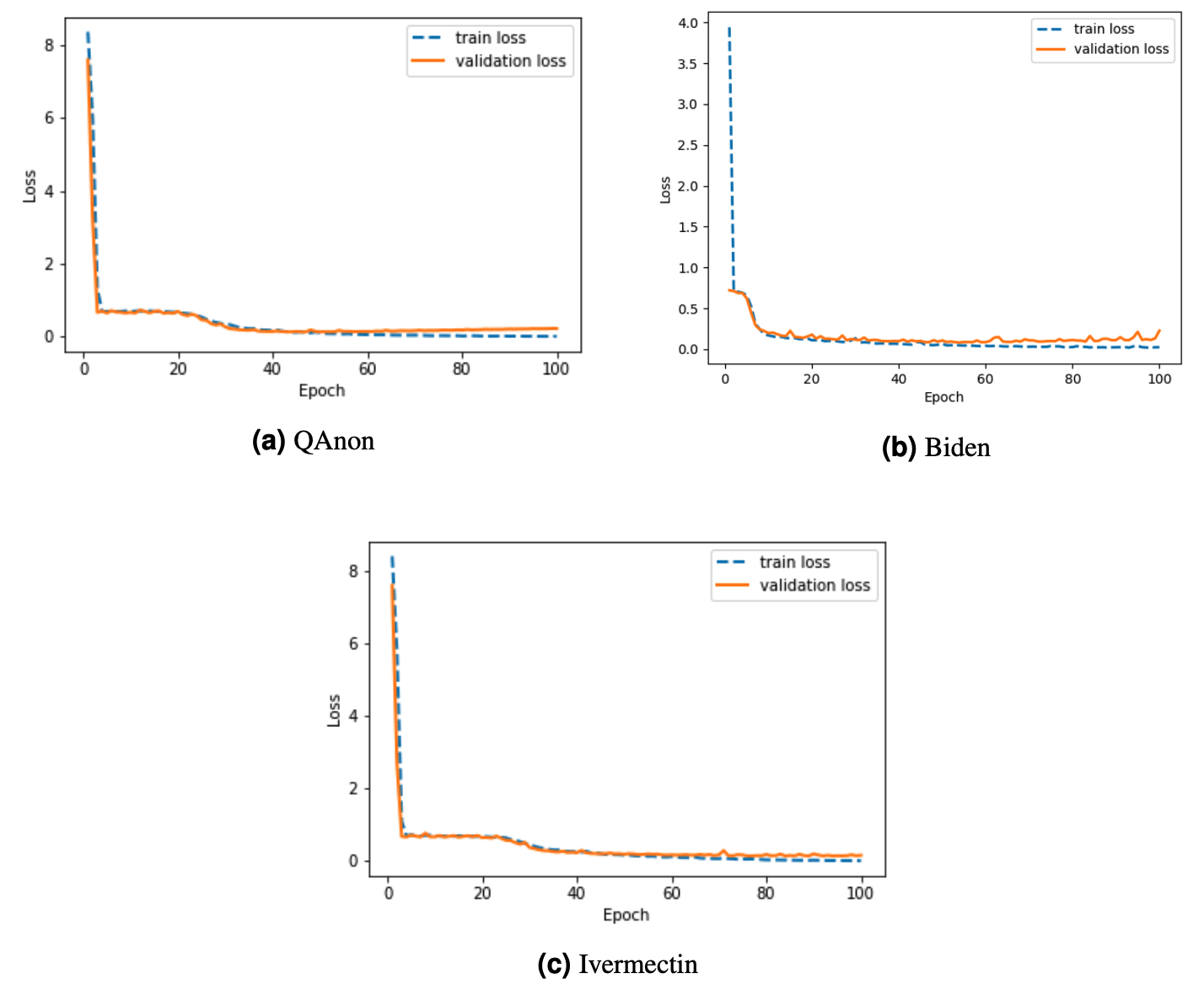}
\end{center}
\caption{Learning curves of selected topics with the domain-based user embedding  for the topics of (a) QAnon, (b) Biden, and (c) Ivermectin.}
\label{fig:learning_curve}
\end{figure}

We trained our model (i.e., domain-based user embedding) and tracked the training and validation loss. 
%Figure~\ref{fig:learning_curve} shows the learning curves where the training loss and validation loss both gradually decreased; that outcome indicates that our model was successful in learning across the three topics. 
The learning curves in Figure~\ref{fig:learning_curve} illustrate the training dynamics of our domain-based embedding model. For QAnon and Biden, the training and validation losses converge smoothly by epoch 50, stabilizing at approximately .15 and .18, respectively. This indicates robust model convergence and good generalization. The smooth curves suggest stable training, and the close alignment between training and validation losses confirms the model's generalization ability. In contrast, for Ivermectin, a slight divergence between training (.12) and validation (.22) losses after epoch 40 suggests reduced generalization, likely due to the smaller dataset size (23,047 users; see Table 3). This clarifies the implications of the learning curves.

%Figure 4

Next, we evaluated our method using validation data and test data using accuracy, F1 score, precision, recall, and ROC-AUC score.
Table~\ref{tab:results} shows the performance of our model with the benchmark models on the validation and test dataset. Overall, the domain-based user embedding performed well across multiple metrics, with the only exceptions being test accuracy and test precision for the topic of Ivermectin, which were nearly comparable to other results. In particular, our model achieved higher test accuracies of .95, .92, and .72 and scored the highest macro-F1 scores of .83, .79, and .52 for the topics of QAnon, Biden, and Ivermectin, respectively. Therefore, in response to RQ1 and RQ2, our results suggest that domain co-occurrence in information sharing is an effective indicator for identifying user clusters during competing events.

\begin{landscape}
\begin{table}[h]
% \begin{adjustwidth}{-4.3cm}{}
\centering
\caption{Performance of user classification across different models.}
\begin{tabular}{ccccccccc}
\hline
Topic                       & Embedding Feature            & Val. Acc. & Test Acc. & Val. macro-F1 & Test macro-F1 & Test presicision & Test recall & Test ROC-AUC \\ \hline
\multirow{4}{*}{QAnon}      & Tweet text         & 0.67      & 0.65      & 0.53.         & 0.48          & 0.47        & 0.5    & 0.5     \\ \cline{2-9} 
                            & Retweet network    & 0.5       & 0.54      & 0.47          & 0.49          & 0.49        & 0.5    & 0.5     \\ \cline{2-9} 
                            & Retweeted articles & 0.56      & 0.64      & \textbf{0.42} & 0.5           & 0.67        & 0.63   & 0.63    \\ \cline{2-9} 
                            & Hashtag-based      & 0.82      & 0.85      & \textbf{0.56} & 0.66          & 0.55        & 0.46   & 0.46    \\ \cline{2-9}
                            & Domain-based       & 0.90      & 0.95      & \textbf{0.70} & 0.83          & 0.90        & 0.87   & 0.87    \\ \hline
\multirow{5}{*}{Biden}      & Tweet text         & 0.57      & 0.57      & 0.46          & 0.45          & 0.47        & 0.5    & 0.5     \\ \cline{2-9} 
                            & Retweet network    & 0.5       & 0.54      & 0.48          & 0.49          & 0.46        & 0.5    & 0.5     \\ \cline{2-9} 
                            & Retweeted articles & 0.64      & 0.69      & \textbf{0.49} & 0.52          & 0.57        & 0.48   & 0.48    \\ \cline{2-9} 
                            & Hashtag-based      & 0.79      & 0.73      & 0.53          & 0.5           & 0.67        & 0.64   & 0.64    \\ \cline{2-9} 
                            & Domain-based       & 0.93      & 0.92      & \textbf{0.79} & 0.79          & 0.72        & 0.69   & 0.69    \\ \hline
\multirow{5}{*}{Ivermectin} & Tweet text         & 0.53      & 0.55      & \textbf{0.46} & 0.41          & 0.57        & 0.56   & 0.56    \\ \cline{2-9} 
                            & Retweet networks   & 0.5       & 0.54      & 0.48          & 0.49          & 0.48        & 0.5    & 0.5     \\ \cline{2-9} 
                            & Retweeted articles & 0.79      & 0.74      & 0.54          & 0.5           & 0.62        & 0.57   & 0.57    \\ \cline{2-9} 
                            & Hashtag-based      & N/A         & N/A         & N/A             & N/A             & N/A           & N/A      & N/A       \\ \cline{2-9} 
                            & Domain-based       & 0.9       & 0.72      & \textbf{0.7}  & 0.52          & 0.61        & 0.64   & 0.64    \\ \hline
\end{tabular}
\label{tab:results}
% \end{adjustwidth}
\end{table}
\end{landscape}

To elucidate the superior performance of our model, we analyze the metrics presented in Table~\ref{tab:results}. Because users tend to share information from websites that align with their beliefs, domain-sharing behavior serves as a strong proxy for user identity in polarized settings. For QAnon, our method achieves a macro-F1 score of $.70$, outperforming the retweet ($.42$) and hashtag embeddings ($.56$). This improvement reflects our model's ability to capture domain-sharing contrasts: pro-QAnon users often share fringe sites like \url{rt.com}, while anti-QAnon users favor mainstream sources like \url{cnn.com}. For Biden, the macro-F1 score is $.79$, $30$ points above the retweet baseline ($.49$). Domain preferences again align with political identity—Democrats with \url{nytimes.com}, Republicans with \url{foxnews.com}, which our method captures better than text-based models, where both sides use overlapping terms (e.g., ``Trump'' or ``election''). For Ivermectin, despite fewer users and no usable hashtags, our model still scores $.70$, outperforming the text-based baseline ($.46$). This shows that domain co-occurrence provides a stable signal even when other features are sparse or noisy.

For the topic of Ivermectin, we could not obtain enough samples to construct hashtag-based user embeddings and therefore could not include them. This limitation demonstrates a key advantage of our domain-based approach: while hashtag usage varies dramatically across topics (from 1,000 users for Biden to only 18 for Ivermectin), domain sharing remains more consistent, enabling reliable user embedding even for topics with limited hashtag engagement. That outcome suggests that hashtag-based user embeddings may be ineffective for topics with few co-occurring hashtags. Adding hashtags to posts or reposts involves manual effort (i.e., typing hashtags), which is inherently costly. By contrast, sharing information via URLs (i.e., domains) is less burdensome and leads to more frequent domain co-occurrence, thereby making it better suited for constructing user embeddings. Furthermore, our results indicate that relying solely on the retweet network may be insufficient for user classification. We expected that text-based information might be effective for user classification in competing events given empirical evidence in previous research~\cite{de2014user,info:doi/10.2196/publichealth.8060,Pennacchiotti_Popescu_2021}.
However, during the linguistic feature evaluation extracted from text and retweeted online articles, the overall performance of the benchmark models was much worse than our model's. Retweeted online articles on Ivermectin reached an accuracy of .74, slightly higher than that of the domain-based user embedding (.72), but its running time of 3,487 s was 60 times longer than the latter's (i.e., 57 s). The lower efficiency of the baseline model might be ascribed to the immensity of the parameters in the language model, thereby indicating that a transformer structure might not always be an efficient component in an attention-based algorithm. 

\subsection{Application of the proposed method}
A key application of the domain-based embedding is creating two-dimensional visualizations, which allows a geometric understanding of user similarities---in our model, preferences for information sources---as physical proximity within a 2D scatter plot. We applied the technique to visualize user embeddings for the topics of QAnon, Biden, and Ivermectin. Using t-SNE, we transformed low-dimensional network representations, obtained through various embedding methods, into two-dimensional spaces in which each user is depicted as a point. Based on Louvain clustering (see Section \ref{subsec:louvain}), users were categorized into two groups and represented with two distinct colors. Ideally, points of the same color (i.e., indicating similar users) should cluster closely, while points of different colors should remain distinct without overlapping.

\begin{figure}[t!]
\begin{center}
\includegraphics[width=1.0\linewidth]{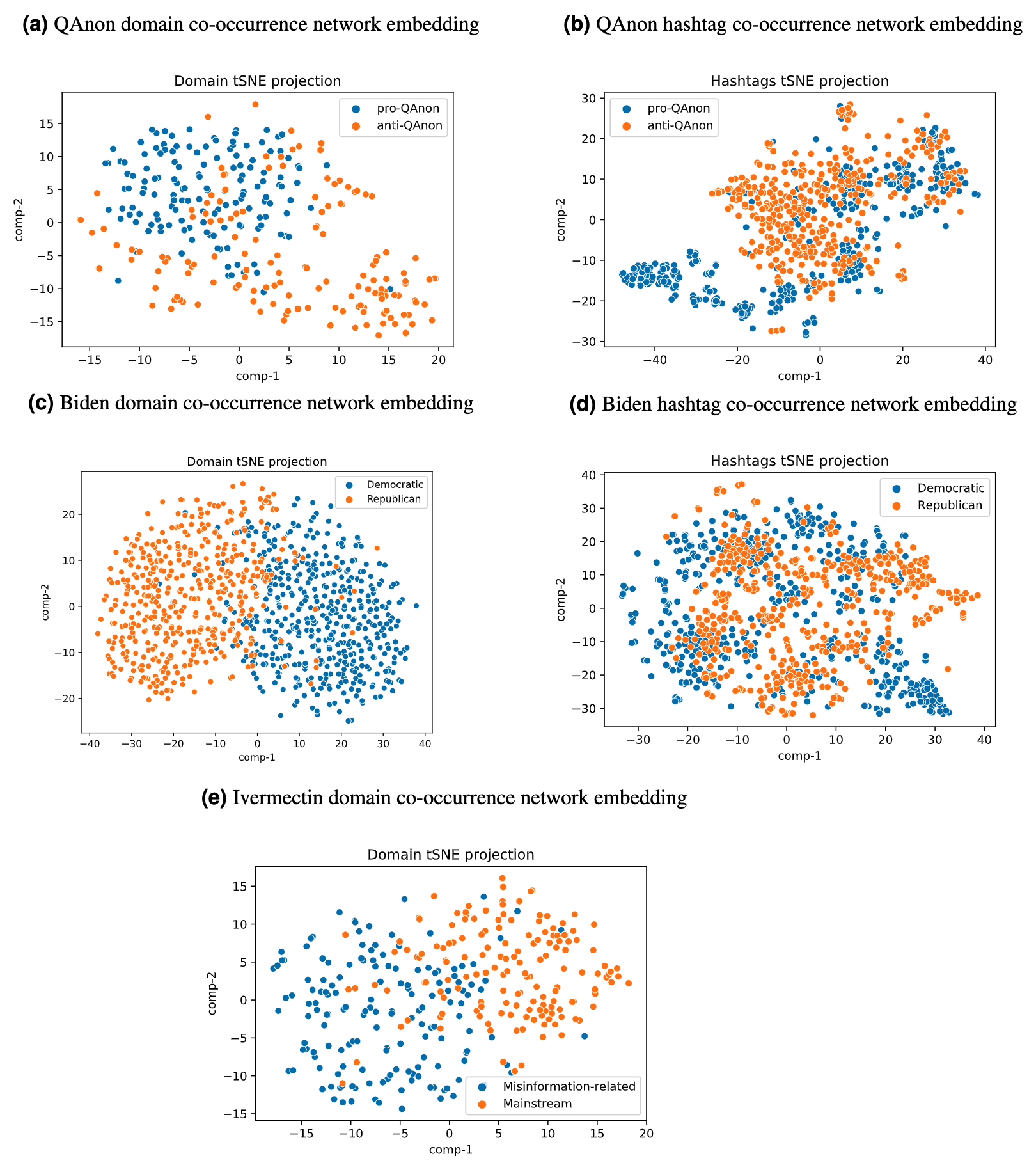}
\end{center}
\caption{t-SNE visualizations of domain-based and hashtag-based user embeddings: (a) and (b) for QAnon topic; (c) and (d) for Biden topic; (e) for Ivermectin topic (Hashtag-based user embedding is not available as explained in the text).}
\label{fig:tsne}
\end{figure}

Figure~\ref{fig:tsne} depicts the graphical representation of domain-based embedding and hashtag embedding. This scatter plot shows that boundaries separating the two classes (i.e., blue and orange) with the domain-based user embedding are much clearer than other methods for the topics of Biden and QAnon than Ivermectin. As previous study showed \cite{Xu2022}, a hashtag co-occurrence network can represent topic relations; therefore, we expected that the latent feature of a hashtag co-occurrence network might help to classify users who support competing topics. However, the results of hashtag-based user embeddings were unsatisfactory because points from the two categories were intermixed. The visualization of domain-based co-occurrence network embeddings is superior in terms of both group separation and border characteristics. 

To quantify visual cluster separation in Figure~\ref{fig:tsne}, we computed silhouette scores, which measure how distinctly each user belongs to its assigned cluster (higher is better). For QAnon (Figure \ref{fig:tsne}a), the visualization indicates well-separated clusters, aligning with the high F1 score of 0.85 in Table~\ref{tab:results} and a silhouette score of 0.39, reflecting effective differentiation of polarized user groups. The Biden topic (Figure ~\ref{fig:tsne}b) shows robust but slightly less distinct clustering, consistent with its F1 score of $0.82$ and a silhouette score of 0.35. In contrast, the Ivermectin plot (Figure ~\ref{fig:tsne}c) shows significant overlap, with a silhouette score of 0.15, reflecting substantial overlap and ambiguity in the embedding space—consistent with the lower F1 score of 0.65 in Table~\ref{tab:results}. This reduced performance is likely due to the smaller dataset size, which results in sparser domain co-occurrence networks, limiting the model's ability to form distinct clusters. These quantitative insights confirm that our method excels in topics with larger datasets, while data sparsity poses challenges to smaller datasets like Ivermectin.

%The findings shown in Table~\ref{tab:results} quantitatively reflect our method's advantage in the user classification. Clusters of the two categories were constructed for Ivermectin (Figure \ref{fig:tsne}c); in the middle, however, several users from the two categories are still jumbled together, with the majority on the right being mainstream users and those on the left being misinformation-related users. Because few hashtags were harvested, hashtag co-occurrence networks could not be produced for Ivermectin. Overall, the t-SNE visualizations confirm that domain-based user embeddings yield much clearer clusters than those generated through hashtag co-occurrence networks.

%Figure 5

\section*{Discussion}
Our method could work on different topics with user networks of different sizes of orders of magnitude. For QAnon and Biden, both the test accuracy and macro-F1 of the proposed method were much higher than those of text and retweeted online article embeddings. Those results could be explained by the possibility that the QAnon conspiracy theory and the 2020 U.S. presidential election could share many words related to the former U.S. president Donald Trump, which might lower the distinguishability of users based on linguistic features. It is common for articles from mainstream domains and fake news domains to share many words, but their underlying stances differ. Additionally, linguistic features in retweeted online articles should not be overlooked, as the test accuracy for retweeted online articles was higher than that for text in the cases of Biden and Ivermectin. Therefore, in some cases, it may be necessary to incorporate appropriate linguistic features before proceeding to further steps to achieve greater accuracy. 

Performance differences across topics are partly explained by dataset size. For example, in the Ivermectin dataset, only 18 users could be embedded using hashtags, compared to 312 using domain co-occurrence (Table 2). This stark contrast reflects that users share URLs more consistently than hashtags. In other words, URLs provide more informative signals than hashtags for user clustering in online competing events. While sparse hashtag usage renders certain methods unusable, our domain-based embedding remains viable and achieves reasonable performance (F1 = .52). This highlights the robustness and practical utility of our method under data sparsity.

Although we used each user's label (e.g., pro-QAnon vs. anti-QAnon) created based on the Louvain method~\cite{blondel2008fast}---if we were to feed the retweet networks as is to Node2vec to obtain embedding representations directly, then the resulting classification accuracy would be far worse than that of the other benchmark models (see Table~\ref{tab:results}). The poor performance could be due to the mechanism that the biased random walk sampling mechanism of Node2vec considers only the network topology structure and excludes the node attribute information of the retweet network.
However, in our method, the likelihood of a mainstream news domain co-occurring with a fake news domain was relatively low, whereas mainstream (or fake) news domains were more likely to co-occur with each other. Therefore, domain co-occurrence may serve as a better proxy for user preferences in competing events, providing more consistent signals even when other behavioral indicators (e.g., hashtags) are sparse.

In a hashtag co-occurrence network, users of both sides can be represented by similar co-occurring hashtags, because it is common for users to attach as many hashtags as they like from both competing events than from one event and not the other. Although the hashtag distributions from both competing events might depend on the topics, there were not enough hashtags to train a classification model for the topic of Ivermectin, and it is therefore not shown in Figure~\ref{fig:tsne}.
The proposed method also has the merit of making it possible to directly compare user similarity with domain-based user embeddings.

Our domain co-occurrence network offers a simpler and more computationally efficient alternative to language models for user embeddings. 
%Indeed, language models such as BERT advance pretrained models to deal with downstream tasks. BERT is trained on 2.5 billion words from the English version of Wikipedia and another 800 million words from BooksCorpus~\cite{1810.04805}.
In our study, Longformer performaned significatly worse with both text and retweeted online articles compared to domain-based user embeddings. Given the high computational cost and complexity of fine-tuning language models, domain co-occurrence networks present a practical alternative.  Moreover, our method enables the comparison of user similarity within clusters. This allows for measuring the similarity between candidate users and malicious users, which can inform the development of recommender systems that promote helpful information. 
%The Ivermectin dataset is smaller than the other two, and the limited number of hashtags prevented us from constructing a hashtag co-occurrence network. Nevertheless, domain-based user embeddings remained viable for user classification. Our model is expected to be applicable in a wide range of scenarios involving social division and polarization.

The variation in dataset sizes, as shown in Table~\ref{tab:data}, significantly impacts model performance. The Ivermectin dataset is substantially smaller than QAnon's users, contributing to its lower F1 score of .65 compared to .85 for QAnon (Table~\ref{tab:results}). Smaller datasets lead to sparser domain co-occurrence networks, reducing the model's ability to capture robust user embeddings, as evidenced by the lower silhouette score of .15 for Ivermectin's t-SNE plot (Figure~\ref{fig:tsne}c). 
In contrast, larger datasets like QAnon and Biden enable denser networks, facilitating more distinct user clusters and higher classification accuracy. This data sparsity effect underscores the importance of sufficient user data for effective domain-based embedding, suggesting that future work should explore techniques like data augmentation to mitigate limitations in smaller datasets.

\section{Conclusion}
This study presented a novel domain-based user embedding method for analyzing competing events on social media, leveraging URL domain co-occurrence patterns to characterize user clusters. This approach outperformed traditional network-based and content-based methods in terms of classification accuracy and computational efficiency. The method was validated using datasets on QAnon, Biden, and Ivermectin, demonstrating its effectiveness in identifying polarized user groups. The results indicate that domain co-occurrence is a robust signal for user clustering, providing deeper insights into user dynamics. In addition, domain-based embeddings also offer practical applications in visualizing user similarities. 
Therefore, our method can be extended to various scenarios related to social divide and polarization, such as echo chambers and dualistic antagonism. 

\section{Future Directions}
Our findings also have limitations. We tested our method on only three relatively large competing events. More nuanced topics need to be studied, however, even if collecting such tweets has become challenging, though not impossible, after the discontinuation of the Twitter Academic API. 
Our method may also be less effective for topics with an insufficient number of available URLs. 
Furthermore, with the recent advancement of large language models (LLMs)~\cite{Minaee2024} and efficient feature learning frameworks~\cite{Liu2024}, LLMs can now be used for user embedding. In such cases, our proposed method can complement these technologies by providing strong features for LLM training and retrieval-augmented generation (RAG), ultimately leading to improve user embedding. Future research should explore more nuanced topics and integrate LLM advancements to further enhance user embedding techniques in computational social science.

\section*{Conflict of Interest}
On behalf of all authors, the corresponding author states that there is no conflict of interest.

\section*{Data Availability Statement}
We confirm that all data analysed during this study are included in this published article: 10.1007/s42001-021-00139-3

\bibliographystyle{spmpsci}
\bibliography{main}

\end{document}